\documentclass[aps,prd,superscriptaddress,11pt,showpacs,notitlepage]{revtex4}
	\usepackage{graphicx}
	\usepackage{amsmath,amssymb,mathrsfs}
	\usepackage[colorlinks=true, a4paper=true, pdfstartview=FitV,
linkcolor=blue, citecolor=blue, urlcolor=blue]{hyperref}

\usepackage{epsf}
\usepackage{epsfig}

\newcommand{\be}{\begin{equation}}
\newcommand{\ee}{\end{equation}}
\newcommand{\bea}{\begin{eqnarray}}
\newcommand{\eea}{\end{eqnarray}}

\newcommand{\bb}{\bibitem}
 
\newcommand{\eqn}{\begin{eqnarray}}
\newcommand{\eqnx}{\end{eqnarray}}
\newcommand{\MM}{\mathcal{M}}
\newcommand{\DD}{\mathbb{D}}
\newcommand{\R}{\mathbb{R}}
\newcommand{\C}{\mathbb{C}}
\renewcommand{\S}{\mathbb{S}}
\renewcommand{\ol}{\overline}
\newcommand{\less}{\backslash}
\newcommand{\ra}{\rightarrow}
\newcommand{\cd}{\partial}
\newcommand{\id}{\rm id}

\numberwithin{equation}{section}

\begin{document}
\title{The volume of a vortex  and the Bradlow bound}
\author{C. Adam}
\affiliation{Departamento de F\'isica de Part\'iculas, Universidad de Santiago de Compostela and Instituto Galego de F\'isica de Altas Enerxias (IGFAE) E-15782 Santiago de Compostela, Spain}
\author{J.M. Speight}
\affiliation{School of Mathematics, University of Leeds, Leeds LS2 9JT, United Kingdom}
\author{A. Wereszczynski}
\affiliation{Institute of Physics,  Jagiellonian University,
Lojasiewicza 11, Krak\'{o}w, Poland}

\begin{abstract}
We demonstrate that the geometric volume of a soliton coincides with the thermodynamical volume also for field theories with higher-dimensional vacuum manifolds (e.g., for gauged scalar field theories supporting vortices or monopoles). We apply this observation to understand Bradlow type bounds for general abelian gauge theories supporting vortices. In the case of {\it SDiff} BPS models (being examples of perfect fluid models) we show that the geometric ``volume" (area) of the vortex, which is base-space independent, is exactly equal to the Bradlow volume (a minimal volume for which a BPS soliton solution exists). This can be finite for compactons or infinite for infinitely extended solitons (in flat Minkowski space-time). 
\end{abstract}

\maketitle 
\section{Introduction}


The probably best-known field theory supporting vortices of finite energy is the abelian Higgs model \cite{Higgs1964}, whose vortex solutions have  been analysed in \cite{Nielsen1973}. The static sector coincides with the Ginzburg-Landau theory of superconductivity \cite{GL1950}, where the corresponding vortex solutions are known as Abrikosov vortices \cite{Abrik1957}. Vortices also play a role as magnetic domains in magnetic materials and may show up in Bose condensates and superfluidity, and they make their appearence in cosmology in the form of co-dimension two defects (cosmic strings), demonstrating both their ubiquitous character and their relevance. 

In this paper, two properties of vortices will be of special importance. The first one is the possibility in some particular field theories to reduce the static field equations (Euler-Lagrange equations) to first-order equations (the so-called BPS equations) \cite{Bogom1976}, \cite{Prasad1975} such that the resulting BPS solutions saturate a global (topological) energy bound. Physically, this means that a multi-vortex BPS configuration (with ``topological index" equal to $N$, say) may be interpreted as a collection of $N$ basic BPS vortices such that the forces between these basic vortices are exactly zero. The second issue is related to the macroscopic or ``thermodynamical" behaviour of multi-vortex configurations, in general (i.e., not necessarily BPS ones). Indeed, a multi-vortex configuration may, e.g.,  behave like a gas of (interacting or non-interacting) vortices, or like a fluid. To study these questions and to calculate average (thermodynamical) quantities, a well-defined notion of the ``volume" (area for the particular case of vortices) of a multi-soliton (multi-vortex) is required. It is one of the main objectives of the present paper to provide this notion and to study its consequences. In the case of vortices, a related feature is the so-called Bradlow bound \cite{Bradlow}, i.e., a minimum ``volume" (area) which a compact base space manifold must have in order to support BPS vortices.

Concretely, in Section II, we provide our definition for the volume of a soliton, generalising the results of \cite{volume}. In Section III, we briefly review the Bradlow bound and zero-temperature thermodynamics (fluid dynamics) of the standard abelian Higgs model, and then study the same issues for certain generalised abelian Higgs models. In Section IV, we repeat this analysis for field theories of the perfect fluid type and, in Section V, for their gauged versions ({\em SDiff} abelian vortices). In Section VI, we briefly discuss the case of conformal solitons. Finally, Section VII contains our conclusions.

The field theories we consider are always defined on a space-time $\mathbb{R} \times \mathcal{M}$ with metric
\be
ds^2 = dt^2 - g_{ij}dx^i dx^j
\ee
for a (positive definite) Riemannian metric $g_{ij}$. Further, the kinetic energy expressions of all our field theories are quadratic in momenta such that, in the static case (which is the case of interest here), the (non-negative) energy density is just minus the static lagrangian density, $\mathcal{E} = -\mathcal{L}$, and the strain tensor is
\be
T^{ij} (x)= - \frac{2}{\sqrt{g}}\frac{\delta}{\delta g_{ij}(x)} E
\ee
where $E = \int d^d x \sqrt{g} \mathcal{E}$ is the energy and $g= {\rm det}g_{ij}$ is the determinant of the metric. Further, we always assume that some units of length and energy have been fixed, such that all constants appearing in the field equations or in the soliton solutions are dimensionless.

In principle, we have to distinguish four different notions of volume in the considerations which follow. The first one is the volume $V_\mathcal{M}$ of the space manifold $\mathcal{M}$. The second is the Bradlow volume $V_B$ (the minimum volume which a space manifold must have such that it supports BPS solutions of a given field theory; obviously, $V_B$ depends on the field theory under consideration). The third volume is the geometric volume $V_{\rm g}$ of the soliton, i.e., the volume of the region where its energy density is nonzero. The fourth volume is the thermodynamic volume $V_{\rm t}$, defined by the (zero temperature) thermodynamic relation
\be \label{thermo-rel}
P_{\rm t}=-\frac{dE}{dV_{\rm t}}
\ee
where $E$ and $P_{\rm t}$ are the energy and the (thermodynamical) pressure. It will turn out, however, that $V_{\rm g}$ and $V_{\rm t}$ may always naturally be assumed to be equal, therefore we use the common symbol $V=V_{\rm g} =V_{\rm t}$. In principle, we also have to distinguish two notions of pressure, namely the thermodynamical pressure $P_{\rm t}$ defined by the relation (\ref{thermo-rel}) and the field theoretic pressure
\be \label{P-f}
P_{\rm f} = \frac{1}{V_{\rm g}} \int d^d x \sqrt{g} \mathcal{P} \; , \quad \mathcal{P} = \frac{1}{d} g_{ij} T^{ij},
\ee
where $\mathcal{P}$ is the pressure density. We will find, however, that precisely when $V=V_{\rm g} =V_{\rm t}$ is assumed, we find as a result that the two pressures are equal and we may use the common symbol $P=P_{\rm t} = P_{\rm f}$. This is discussed in general in Section II, whereas particular examples are considered in the subsequent sections.

\section{Volume and pressure of a soliton}
In \cite{volume} it was proven that $V_{\rm g} =V_{\rm t} \; \Rightarrow \; P_{\rm t} = P_{\rm f}$ for scalar field theories with zero-dimensional vacuum manifolds in flat space $\mathbb{R}^d$. 
Here we want to generalize these results to more general field theories (e.g., gauge theories, like the abelian Higgs model, which supports vortices with a vacuum manifold $\mathbb{S}^1$) and to more general space manifolds. For this purpose, let us consider the following constrained energy functional
\be \label{constr-func}
E_{\rm c} = E + C_{\rm vol} \; , \quad C_{\rm vol} = P\left( \int d^d x \sqrt{g}\, \Theta (\mathcal{E}(x))  -V\right).
\ee
Here, $C_{\rm vol}$ is the volume constraint imposing the condition $V=V_{\rm g}$, and $P$ is the corresponding Lagrange multiplier. Further, the generalised step function $\Theta (\mathcal{E}(x))$ is the locus function of a field configuration, i.e., 
\be \label{gen-step}
 \Theta (\mathcal{E}(x)) =
\left\{
\begin{array}{c}
1 \quad \mbox{for} \quad \mathcal{E}(x) >0 \\
0 \quad \mbox{for} \quad  \mathcal{E}(x) =0
\end{array}
\right.\, 
\ee
which, obviously, implies $\int d^d x \sqrt{g}\, \Theta (\mathcal{E}(x)) = V_{\rm g}$. The explicit expression for the locus function chosen here is different from the one in \cite{volume} (although they coincide for the cases considered in \cite{volume}) and the expression used here allows to consider more general cases. The important point for us is that the locus function is invariant w.r.t. infinitesimal variations of the metric (the relevance of this condition will become clear in a moment). 
This invariance follows from the following observations. Some terms in $\mathcal{E}$ (e.g., potentials) do not depend on the metric, at all, so invariance is obvious. The metric dependent terms in $\mathcal{E}$ may always be written as non-degenerate quadratic forms $v^a M^{ab}v^b$ where $M^{ab}$ is a positive definite matrix constructed from the metric, whereas the $v^a$ are constructed from the fields and their (partial or gauge-covariant) derivatives. Examples are the standard (gauge) kinetic term where $M^{ab} \to g^{ij}$ and $v^a \to \partial_i \phi $ [or $v^a \to D_ i \phi = (\partial_i -ieA_ i) \phi  $ for a complex $\phi$ coupled to a gauge field], or Skyrme-type terms, where $M^{ab} \to g^{ij} g^{kl}$ and $v^a \to \partial_i \phi^1 \partial_k \phi^2 - \partial_ i \phi^2 \partial_k \phi^1$ for two fields $\phi^1$, $\phi^2$. The important point is that all these expressions are zero if and only if  $v^a =0$ (a vacuum configuration), independently of the metric, because of the positive definite nature of $M^{ab}$. It follows immediately that the locus function is invariant under variations of the metric.

The Lagrange multiplier $P$ obeys the thermodynamical relation $P=-(dE_{\rm c}/dV)$, by construction. We now want to prove that it is, at the same time, equal to the field theoretical pressure (\ref{P-f}). In a first step, we prove it for flat space $g_{ij} =\delta_{ij}$, $g=1$, closely following \cite{volume}. We act with an infinitesimal scale transformation $x^i \to (1+\lambda) x^i$, for infinitesimal $\lambda$, on all fields which appear in the constrained energy functional (but not on the metric). Then we assume that all terms in the energy density are ``geometrically natural" \cite{sp1} such that a coordinate transformation acting on the fields may be traded for the inverse coordinate transformation of the metric, $\delta_{ij} \to (1-2\lambda) \delta_{ij}$. That is to say, we act with the transformation
\be
\delta  = \left. \int d^d x \, \delta g_{ij}(\vec x) \, \frac{\delta }{\delta g_{ij}(\vec x)}\right|_{g_{ij}=-\delta_{ij}}
\ee
where $\delta g_{ij} = -2\lambda g_{ij}$ on $E_{\rm c}$. 
Variation of $E$ just gives
\be \label{delta-E}
\delta E = \lambda \int d^d x \sqrt{g} g_{ij}T^{ij} = \lambda d\int d^d x \sqrt{g} \mathcal{P} = \lambda d \, V P_{\rm f}
\ee
whereas variation of $C_{\rm vol}$ gives
\be \label{delta-C}
\delta C_{\rm vol} = - P \lambda d \int d^d x \sqrt{g} \, \Theta(\mathcal{E}) = -P\lambda d\, V
\ee
where we used the invariance of $\Theta (\mathcal{E})$. Further, we wrote the expressions for general $g_{ij}$, for later convenience.  To close the argument, we remember that the variation $\delta$ came from a variation of the fields (a scale transformation). But for solutions of the variational problem (\ref{constr-func}), the functional is invariant under {\em any} field variation, that is, on-shell (for solutions) $\delta E_{\rm c} =0$,  immediately leading to $P=P_{\rm f}$, which is what we wanted to prove. 

The question is whether this argument can be generalised for general space manifolds $\mathcal{M}$. After all, in Eqs. (\ref{delta-E}), (\ref{delta-C}) only the constant Weyl transformation $g_{ij} \to {\rm e}^{-2\lambda}g_{ij}$ appears, which can be defined for any metric. The apparent problem is that, as this Weyl transformation no longer follows from a variation of the fields (i.e., from a coordinate transformation), there is no obvious reason why $\delta E_{\rm c} =0$ should hold on-shell. Nevertheless, if we accept (i.e., define) that the volume of a soliton is varied via a Weyl transformation and not by squeezing the soliton for a fixed metric (on a fixed background manifold), then the result $P_{\rm t} = P_{\rm f}$ follows immediately and does not even require the introduction of the constrained energy functional $E_{\rm c}$. The reason is that in this case the operations of a constant Weyl transformation acting on $E$ (which defines $P_{\rm f}$) and of a variation of the volume (which defines $P_{\rm t}$) are identical. 

To squeeze a soliton (i.e., to introduce pressure) via a Weyl rescaling (i.e., by squeezing the whole manifold) may appear counter-intuitive in a first instant, but here we want to argue that it is a natural definition in the general case. First of all, on $\mathbb{R}^d$ it is equivalent to the standard idea of squeezing the soliton itself. Secondly, for general $g_{ij}$ there are infinitely many possibilities to squeeze a soliton on a fixed background manifold, and there is no preferred ``symmetric" or ``shape-preserving" way of squeezing - unless the metric itself has some symmetries (Killing vectors). Squeezing via a constant Weyl rescaling, on the other hand, is shape-preserving by construction and, therefore, the most ``symmetric" definition to squeeze a soliton, i.e., to change its volume. In addition, if a soliton on a compact manifold covers the whole manifold (i.e., $V=V_\mathcal{M}$), it is not clear how to squeeze the soliton to a smaller volume for a fixed background without drastically changing its boundary conditions. \\
We remark that the Weyl rescaling definition for the variation of the volume of a soliton implies that the pressure of a BPS soliton is always zero. Indeed, the energy of a BPS soliton is proportional to a topological invariant (winding number, etc.) which is metric independent by definition. It follows that the Weyl variation of the on-shell energy and, therefore, the pressure, is identically zero.

\section{The Bradlow bound - Abelian Higgs Vortices}
\subsection{The Abelian Higgs Vortices at critical coupling}
After having established that the geometric volume is, in fact, the thermodynamic volume, we aim to investigate its relation with the Bradlow volume $V_B$ and the volume of the space manifold $V_\mathcal{M}$. Furthermore, we show that the Bradlow volume may be used as a rather universal characteristic of general BPS type theories. 

Let us briefly demonstrate the original derivation of the Bradlow bound for solitons in the Abelian Higgs model at critical coupling on an arbitrary compact manifold $\mathcal{M}$ \cite{Bradlow}, \cite{MS}
\be
L=\int_\mathcal{M} \left( -\frac{1}{4} F_{\mu \nu}^2+\frac{1}{2} \overline{D_\mu u}D^\mu u -\frac{1}{8} (1-u\bar{u})^2\right) \Omega d^2x
\ee
where $D_\mu u = (\partial_\mu -iA_\mu) u$, and the metric is
\be \label{met-Omega}
ds^2=dt^2-\Omega (x^1, x^2)\left( (dx^1)^2+(dx^2)^2\right) .
\ee
The energy bound is
\be
E \geq \pi N
\ee
where $N$ is the topological charge, and the static energy is
\be
E=\frac{1}{2} \int_\mathcal{M} \left( \Omega^{-1} B^2 + \overline{D_1 u} D_1 u + \overline{D_2 u} D_2 u +\frac{\Omega}{4} (1-u\bar{u})^2\right)d^2x .
\ee
The bound is saturated for configurations obeying the Bogomolny equations
\bea
D_1u+iD_2u &=&0 \\
B-\frac{\Omega}{2} (1-u\bar{u})&=&0. \label{stand-BPS2}
\eea
To get the Bradlow bound, we integrate the last equation over the physical (base) space manifold $\mathcal{M}$. Then,
\be
2\int_\mathcal{M} B d^2x + \int_\mathcal{M} u\bar{u} \Omega d^2x  = \int_\mathcal{M} \Omega d^2x.
\ee
If the base space is a compact surface without boundary, the first Chern number must be an integer $N$, i.e., the magnetic flux $\Phi_B \equiv \int d^2 x B = 2\pi N$, so
\be
4\pi N + \int_\mathcal{M} u\bar{u} \Omega d^2x  = V_{\mathcal{M}}
\ee
where $V_{\mathcal{M}}=\int_\mathcal{M} \Omega d^2x$ is the total area (two dimensional volume) of the physical space. 
Hence, finally, we arrive at the bound 
\be
V_{\mathcal{M}} \geq 4\pi N.
\ee
The meaning of the bound is the following. For a given value of the topological charge $N$, solutions of the Bogomolny equations (BPS vortices) do not exist unless the area of the (compact) base space is bigger than $4\pi N$. For a a fixed volume of the base space manifold, one can put a finite number of BPS vortices up to a maximal number (provided by the bound), $N\le (V_\mathcal{M}/4\pi)$, above which vortices are no longer solutions of the Bogomolny equations but, instead, solve the full second order equations of motion.
Here we introduce the notion of the Bradlow volume $V_B$ as the minimal volume of the manifold $\mathcal{M}$ for which BPS vortices (of a given topological charge) do exist. Obviously, this quantity can be defined for any model with BPS solitons.

\vspace*{0.2cm}

{\it Definition.} The Bradlow volume $V_B$ is the minimal volume of the base space (compact) manifold $\mathcal{M}$ for which the BPS sector of a solitonic model has a non-trivial solution. 
\vspace*{0.2cm}

The energy of the BPS vortices ($V_\mathcal{M} \geq V_B$) is independent of the base space volume and takes the constant value 
\be
E=\pi N.
\ee 
Below the bound, for $V_\mathcal{M} < V_B$, there only exists the ``constant" non-BPS solution of the full second-order EL equations 
\be
u=0\, , \quad B= 2\pi N V^{-1} \Omega
\ee
 with energy
\be
E[V]=\frac{(2\pi N)^2}{2V}+\frac{V}{8}.
\ee
In all cases, the geometric (thermodynamical) volume $V$ of a vortex is equal to $V_\mathcal{M}$, as the vortices cover the full base space. Then we define pressure as
\be
P =-\left( \frac{\partial E}{\partial V}\right) .
\ee
For BPS solutions, the energy $E=\pi N$ is independent of the volume, hence $P=0$ for $V \ge V_B$. Note that also the pressure density is identically 0 in this case due to the Bogomolny equations. Below the bound, ($V\leq V_B$) a nontrivial positive pressure appears
\be
P=\frac{(2\pi N)^2}{2V^2}-\frac{1}{8} = \frac{1}{8} \left(  \frac{V_B^2}{V^2} -1\right) .
\ee
As a consequence, one could say that the Abelian-Higgs model on a compact manifold at critical coupling and large (topological charge or energy) density describes a solitonic matter with maximally stiff mean-field equation of state. In fact, the mean field energy density reads
\be
\bar{\epsilon}=\frac{E}{V}=\frac{(2\pi N)^2}{2V^2}+\frac{1}{8}=\frac{1}{8} \left(  \frac{V_B^2}{V^2} +1\right)
\ee
Hence,
\be
\bar{\epsilon}=P +\frac{1}{4}.
\ee
This equation of state for vortices is valid for the case of a compact manifold $\mathcal{M}$ without boundary. This means that it does not have to be identical to the equation of state of vortex matter in the plane, which represents a system of vortices enclosed in a finite volume subspace of the original $\mathbb{R}^2$ space. In such a situation, that is a vortex in a finite flat disc $\mathbb{D} \subset \mathbb{R}^2$, one has to carefully face the issue of the boundary conditions. On $\mathbb{R}^2$, the first Chern number does not have to be an integer, and the quantization of the magnetic flux is, instead, a consequence of the condition of finite energy. In polar coordinates $\mu = t,r,\varphi$, finite static energy requires $\lim_{r\to \infty} u = {\rm e}^{i \alpha (\varphi)}$ and, in the gauge $A_t =0$, $A_r =0$, $\lim_{r\to \infty} A_\varphi = \partial_\varphi \alpha$, and the vortex number is given by the winding of $\alpha$ at infinity, $\alpha (2\pi) -\alpha (0) = 2\pi N$.

On a finite disc $\mathbb{D} \subset \mathbb{R}^2$, there exist two possibilities to put a vortex with vorticity $N$ such that $u$ at the boundary of $\mathbb{D}$ still is $u\vert_{\partial \mathbb{D} } = {\rm e}^{i\alpha}$ and $\alpha$ changes by $2\pi N$ while traversing the boundary once (i.e., imposing Dirichlet boundary conditions on the field $u$).
The first possibility consists in requiring that the vortex still obeys the BPS equations. It is then no longer possible to require that the gauge field approaches a pure gauge at the boundary and, consequently, the magnetic flux is no longer quantised \cite{Nasir}. The easiest way to see this is by assuming a round disc with radius $R$ and the spherically symmetric ansatz
\be \label{spher-sym}
u(r,\varphi) = {\rm e}^{iN\varphi} f(r) \; ,\quad A_\varphi = N-a(r) \quad \Rightarrow \quad B=-\frac{a'}{r}
\ee
leading to the BPS equations
\be
f' = \frac{af}{r} \, , \quad a' = -\frac{1}{2} r(1-f^2).
\ee
On $\mathbb{R}^2$, the boundary conditions for a vortex are $f(0)=0$, $a(0)=N$ and $\lim_{r\to \infty} f(r)=1$, $\lim_{r\to \infty}a(r)=0$.
On a finite disc, the two conditions $f(R)=1$ and $a(R)=0$ together, however, are too strong (as may be checked easily) and only permit the trivial solution $f=1$, $a=0$. Requiring $f(R)=1$ (Dirichlet boundary condition) therefore forces us to allow for a nonzero $a(R) \equiv a_R$, and the magnetic flux $\Phi_B = 2\pi \int dr r B =2\pi (N-a_R)$ is no longer quantised. The energy, on the other hand, still takes the BPS value $E=\pi N$, and the pressure is, therefore, zero. These solutions exist on discs of arbitrary size \cite{Nasir}, therefore there is no Bradlow bound for the Abelian Higgs model at the critical coupling on a flat finite disc $\mathbb{D}$. 
Both the fact that the pressure remains zero and that the magnetic flux changes its value with the disc size implies that these vortices should not be interpreted as ``squeezed" versions of the original vortex on $\mathbb{R}^2$. 

The second possibility consists in requiring both that $u\vert_{\partial \mathbb{D} } = {\rm e}^{i\alpha}$ and that $A_\mu$ is pure gauge at the boundary of the disc, such that the magnetic flux remains quantised. The corresponding vortex can no longer obey the BPS equations, but it may still solve the full (static) second-order Euler-Lagrange equations. 
In particular, for a spherically symmetric disc and for the spherically symmetric ansatz from above, we may now impose $f(R)=1$ and $a(R)=0$. The resulting solution has a quantised magnetic flux and a nonzero pressure (because it is not BPS) and may, therefore, now be interpreted as a squeezed version of the spherically symmetric BPS vortex on $\mathbb{R}^2$.

We remark that in the recent paper \cite{five-vort} several version of vortex equations related to but different from the standard BPS equation (\ref{stand-BPS2}) have been considered, leading to interesting variations of the Bradlow bound.  
In one case, a BPS vortex equation with an ``inverse" Bradlow bound $N\ge (V_\mathcal{M}/4\pi)$ has been found, where the vortex number must be above a certain minimum value. The BPS vortices in this model are, however, critical points but not minima of the energy functional, which is not even bound from below. The corresponding vortex equation itself was already introduced in \cite{Am-Ol} in a different context. Another BPS vortex equation studied in \cite{five-vort} (``Bradlow vortices") may have solutions only if the volume of the manifold is equal to the Bradlow volume, so that the ``Bradlow bound" is converted into a ``Bradlow equation" (for more on this model see \cite{Bja2017}).

\subsection{Generalized Abelian Higgs Vortices}
It has been observed recently \cite{bazeia} that one can generalize the Abelian Higgs model preserving its BPS property by allowing for two field dependent non-negative coupling functions which multiply the gauge and the Higgs kinetic parts of the action
\be
L_{gen}=\int \left( -\frac{G(|u|^2)}{4} F_{\mu \nu}^2+\frac{w(|u|^2)}{2} \overline{D_\mu u}D^\mu u -U \right) \Omega d^2x .
\ee
Here, the potential $U(u\bar u)$ is always assumed to take its vacuum value at $|u|=1$, i.e., $U(1)=0$. Further, 
 $w$ is related to a nontrivial target space metric, whereas $G$ defines a kind of generalised magnetic permeability (we remark that the case $G=1$, $w\not= 1$ was already considered in \cite{Lohe}).
The model has a BPS sector if the coupling functions $G, w$ and the potential $U$ obey ($f\equiv |u|$)
\be
\frac{d}{df} \sqrt{GU}=\frac{1}{\sqrt{2}} w f
\ee
or, for the simplifying assumption $w=1$, 
\be
GU = \frac{1}{8}(1-f^2)^2 .
\ee
The Bogomolny equations in this case are \cite{bazeia}, \cite{contatto}
\bea \label{BOG1}
D_1u+iD_2u &=&0 \\ \label{BOG2}
B-\frac{\Omega}{2} \frac{(1-u\bar{u})}{G(u\bar{u})} \equiv B - 4\Omega \frac{U(u\bar u)}{1-u\bar u}&=&0,
\eea
and the corresponding BPS energy is still $E_{\rm BPS} = \pi N$.
The question is whether for these models, when defined on a compact space manifold $\mathcal{M}$ without boundary, there exists a Bradlow bound $V_B$ such that the above BPS equations may be solved for $V_\mathcal{M} \ge V_B$. Here we want to give some indications that this may be the case, at least for some choices of $G$. First of all, the above BPS equations have the ``constant", trivial solution 
\be \label{const-sol-BPS}
u=0, \quad B= \frac{\Omega}{2G(0)}.
\ee
 As a consequence of the magnetic  flux quantisation, the manifold where this solution exists must have the ``Bradlow" volume
\be \label{brad-vol-G}
V_B = 4\pi NG(0) = \frac{\pi N}{2U(0)}.
\ee
Next we observe that, for arbitrary volume $V_\mathcal{M}$, the full second-order EL equations have the ``constant" solution
\be \label{const-sol}
u=0\, , \quad B = 2\pi NV_\mathcal{M}^{-1} \Omega .
\ee  
The corresponding static energy is
\be
E = \frac{1}{2} \int d^2 x \left( \frac{G(0)}{\Omega} B^2 + 2\Omega U(0) \right) = (2\pi N)^2\frac{G(0)}{2V} + U(0)V
\ee
(where $V=V_\mathcal{M}$).
Further, this energy takes its minimal value $E=E_{\rm BPS}=\pi N$ exactly at $V_\mathcal{M}=V_B = 4\pi NG(0)$ (where we used that $U(0)=(8G(0))^{-1}$) and leads to positive pressure for $V_\mathcal{M}<V_B$ but to negative pressure for $V_\mathcal{M}>V_B$. This indicates an instability of the formal solution (\ref{const-sol}) for $V_\mathcal{M}>V_B$, i.e., the existence of field configurations with lower energy. A different question is whether local extrema (static solutions) or global minima (BPS solutions) of the energy functional may be found among these low energy configurations. A general answer to this question for general $G$ is beyond the scope of the present paper. We will also find some indications that the answer to this question may be quite sensitive to the choice of the coupling function $G$. For, let us assume that BPS solutions exist for $V_\mathcal{M}\ge V_B$ and provide the true, stable solutions. 
A first, related question to ask is under which circumstances the ``Bradlow volume" (\ref{brad-vol-G}) (which has a trivial solution with BPS energy) provides, at the same time, a Bradlow bound (such that BPS solutions can exist only for $V_\mathcal{M}\ge V_B$, but not for $V_\mathcal{M}<V_B$). Integrating the second BPS equation over the manifold, the resulting equation may be expressed like
\be
V_B = V_\mathcal{M} - \frac{1}{U(0)} \int d^2 x \Omega \left( U(0) - \frac{U(u\bar u)}{1-u\bar u}\right) .
\ee
This implies $V_B \le V_\mathcal{M}$ provided that the second term at the r.h.s. is non-negative. A sufficient condition for this is
\be \label{pot-ineq}
U(u\bar u) \le U(0)(1-u\bar u).
\ee
This condition is also necessary in a certain sense, because otherwise the condition $V_B \le V_\mathcal{M}$ would not be universal and would depend on the field configuration $u$ (it would be broken by fields which take such values in most of $\mathcal{M}$ which violate the above inequality) and on the volume element $\Omega$ ($\Omega$ could enhace the contribution of these ``forbidden" regions). A class of potentials which obey (\ref{pot-ineq}) is
\be
U=(1-u\bar u)^\alpha, \quad \alpha \ge 1,
\ee 
corresponding to $G=(1/8)(1-u\bar u)^{2-\alpha}$.

Now, let us assume that we chose a potential obeying (\ref{pot-ineq}) and, further, let us {\em assume} that BPS vortex solutions exist for $V_\mathcal{M} >V_B$. Then these BPS vortex solutions may show two rather different types of behaviour for $V_\mathcal{M}$ close to $V_B$. The first possibility is that the BPS vortex solution approaches the constant solution (\ref{const-sol-BPS}) in the limit $V_\mathcal{M} \to V_B$.  In other words, the stable ``constant" solution (\ref{const-sol}) for $V_\mathcal{M} <V_B$ bifurcates at $V_\mathcal{M} =V_B$ into an unstable constant solution and a stable BPS solution. This is known to happen, e.g., for the standard abelian Higgs model at critical coupling ($U=(1/8)(1-u\bar u)^2$). It is plausible to conjecture that this behaviour will occur for deformation functions $G$ which do not deviate too much from the standard abelian Higgs case $G=1$. In physical terms, one might say that at $V_\mathcal{M} = V_B$ a second order phase transition occurs. The second possibility is that the BPS vortex solution is completely different from the constant solution (\ref{const-sol-BPS}) even for $V_\mathcal{M}$ very close to $V_B$, despite the fact that they have the same energy. In the limit $V_\mathcal{M} \to V_B$ the BPS vortex may either approach a solution different from (\ref{const-sol-BPS}), or the limit is not attainable, i.e., a limiting solution does not exist. In both cases, the phase transition is of first order. This possibility of a first-order phase transition may appear strange at first sight, but we will find that it exists and is realised by compact BPS vortex solutions, i.e., BPS vortices which deviate from their vacuum value only in a finite region with the topology of a disc.  
Two particular cases of compact vortices allowing for explicit solutions are considered in the next subsection. 

We end this section by remarking that there exists the particular case $G = |u|$ which allows to reduce the BPS equations on $\mathbb{R}^2$ to the (integrable) elliptic sinh-Gordon equation for $g=\ln |u|$. It can be demonstrated that in this case a vortex solution exists for $N=1$ but not for higher vortex number \cite{dun}, although an explicit expression for this solution of the elliptic sinh-Gordon equation (obeying the boundary conditions of a vortex) cannot be found. 

\subsection{Compact Generalized Abelian Higgs Vortices}
Very recently, some examples of compact BPS  vortices, i.e., BPS vortices which deviate from their vacuum value only in a finite subregion (the ``locus set" of the vortex) of $\mathbb{R}^2$, have been constructed for the generalized abelian Higgs model in \cite{bazeia2016}. This implies that these compact vortices continue to exist on space manifolds which are themselves finite subsets of $\mathbb{R}^2$ (discs), provided that the locus set of the compact vortex is completely contained within the disc. A different question is whether the corresponding models may support BPS vortices on compact manifolds without boundary. We will find that the characteristic behaviour of such solutions depends on the specific model under consideration, therefore we just consider two particular examples found in \cite{bazeia2016}. We remark that the compacton solutions in \cite{bazeia2016} were constructed as limiting cases of regular, non-compact vortices, demonstrating that compactons (with low regularity at the compacton boundary) may be understood as limiting cases of non-compact vortices with high regularity (infinitely differentiable).

\subsubsection{Case $G=|1-|u|^2|$}
For $G=|1-f^2|$, $f\equiv |u|$, (and $w=1$), the resulting potential reads
\be
U=\frac{1}{8}|1-f^2|.
\ee
This potential approaches the vacuum at $f=1$ less than quadratically (in fact, linearly) and, therefore, leads to compacton solutions, which are of low regularity at the compacton boundary.
Using the spherically symmetric ansatz (\ref{spher-sym}) in $\mathbb{R}^2$, the resulting BPS equations in the fundamental domain of the vortex (for $0\le f\le 1$) are
\be
f' = \frac{af}{r} \; , \quad a' = -\frac{r}{2}.
\ee
The second equation has the solution
\be
a = \left\{
\begin{array}{cc}
 N-\frac{r^2}{4}\,  & \;\; r \leq 2\sqrt{N}
\\
 & \\
 0 & r > 2\sqrt{N} \, ,
\end{array}
\right. 
\ee
and the compacton radius is $r_c = 2\sqrt{N}$, leading to the compacton volume $V_c = \pi r_c^2 = 4\pi N$, which exactly agrees with the Bradlow volume $V_B = 4\pi NG(0)= 4\pi N$. The gauge potential $a$ is continuous at the compacton boundary, but the resulting magnetic field
\be
B = \left\{
\begin{array}{cc}
 \frac{1}{2}\,  & \;\; r \leq 2\sqrt{N}
\\
 & \\
 0 &\; \;  r > 2\sqrt{N} \, 
\end{array}
\right. 
\ee
is discontinuous. Finally, integration of the first BPS equation leads to
\be
f = \left\{
\begin{array}{cc}
\left( \frac{r}{2\sqrt{N}}\right)^N {\rm e}^\frac{2N-r^2}{8} & \;\; r \leq 2\sqrt{N}
\\
 & \\
 1 &\; \;  r > 2\sqrt{N} \, 
\end{array}
\right. 
\ee
Both $f$ and $f'$ are continuous at $r=r_c =2\sqrt{N}$.

The fact that the Bradlow volume and the compacton volume agree makes one suspect that compact BPS vortices continue to exist on compact manifolds $\mathcal{M}$ without boundary provided that the volume of the manifold is sufficiently big, $V_\mathcal{M} > V_B$. Here we shall demonstrate that this is indeed true for the specific case $\mathcal{M} = \mathbb{S}^2$. 
For our purposes, it is useful to generalise the two BPS equations (\ref{BOG1}), (\ref{BOG2}) (which, as they stand, only hold for metrics of the form (\ref{met-Omega})) to space metrics on $\mathcal{M}$ in arbitrary coordinates, i.e., $ds^2_\mathcal{M} = g_{ij} dx^i dx^j$. 
For the second BPS equation this is easy, because it may immediately be written as an equation between two-forms,
\be
{\bf B} = \frac{1}{2} {\bf vol}
\ee
where ${\bf B} = d{\bf A}$ is the magnetic two-form, and ${\bf vol}$ is the ``volume" form (area two-form) on $\mathcal{M}$. In particular, for the metric on $\mathbb{S}^2$ in longitude and latitude coordinates,
\be \label{met-S2}
ds^2_{\mathbb{S}^2} = R^2 (d\theta^2 + \sin^2 \theta d\varphi^2 )
\ee
(where $R$ is the  - dimensionless - radius of the sphere), and for the rotationally symmetric ansatz 
\be \label{rot-ansatz}
{\bf A} =(N-a(\theta))d\varphi \; , \quad u = f(\theta) {\rm e}^{iN\varphi} \, ,
\ee
this equation leads to
\be
a_{,\theta} = -\frac{R^2}{2} \sin \theta
\ee
and the formal solution obeying the boundary condition $a(\theta =0)=N$ is
\be
a = \frac{R^2}{2} (\cos\theta -1) +N.
\ee
This solution may be extended to a compacton solution
\be
a = \left\{ 
\begin{array}{cc}
\frac{R^2}{2} (\cos\theta -1) +N \,  & \;\; \theta \leq \theta_c
\\
 & \\
 0 & \theta > \theta_c \, ,
\end{array}
\right. \qquad \cos \theta_c = \frac{R^2 -2N}{R^2}
\ee
provided that $R^2> N$, that is, $V_{\mathbb{S}^2} \equiv 4\pi R^2 > 4\pi   N \equiv V_B$. So, as expected, the compacton solutions on $\mathbb{S}^2$ exist if the area of the two-sphere exceeds the Bradlow ``volume" (area) $V_B$. 

The first BPS equation (\ref{BOG1}) may be expressed in a coordinate-independent way like follows. Temporarily renaming the coordinates like $x^1 \to x, \; x^2 \to y$, (\ref{BOG1}) reads $D_y u = i D_x u$ or, after introducing the complex-valued one-form
\be
{\rm D}u = D_x u\, dx + D_y u\, dy,
\ee
${\rm D}u (\partial_y) = i {\rm D}u (\partial_x)$. In other words, evaluating ${\rm D}u$ on the vector rotated by 90 degrees in the counter-clockwise sense $\partial_x \to \partial_y$ is equivalent to multiplying ${\rm D}u(\partial_x)$ by $i$. But this has a simple coordinate-invariant generalisation on any oriented Riemannian two manifold $\mathcal{M}$. Then ${\rm D}u$ is a linear map from the tangent space to the complex numbers, ${\rm D}u: T_p\mathcal{M} \to \mathbb{C}$. We may define the following complex structures on the two vector spaces $\mathbb{C}$ and $T_p\mathcal{M}$. On $\mathbb{C}$ it is just multiplication by $i$, and on $T_p \mathcal{M}$, rotation by 90 degrees counterclockwise, as determined by the metric and orientation on $\mathcal{M}$. Then the first BPS equation in its coordinate-invariant form just says that ${\rm D}u$ is complex linear, that is,
\be
{\rm D}u(JX) = i{\rm D}u(X)
\ee
for arbitrary vectors $X$. In particular, for the metric (\ref{met-S2}) and for the vector $X=\partial_\theta \; \Rightarrow \; JX =(1/\sin\theta ) \partial_\varphi$, this equation reads $(1/\sin\theta )D_\varphi u = i D_\theta u$ which, for the ansatz (\ref{rot-ansatz}), leads to the first-order ODE
\be
f{,_\theta} = \frac{af}{\sin\theta} .
\ee
The solution may be found easily by introducing the new variable $t=\cos\theta$ and, 
 with the correct choice of normalisation, reads
\be
f =\left\{ 
\begin{array}{cc}
\left( \frac{\cos \theta +1}{\cos\theta_c +1} \right)^\frac{R^2 -N}{2} \left( \frac{1-\cos\theta}{1-\cos \theta_c}\right)^\frac{N}{2}
  \,  & \;\; \theta \le \theta_c
\\
 & \\
 1 & \theta > \theta_c \, .
\end{array}
\right. 
\ee
It may be checked that $f(\theta_c) =1$, and $f' (\theta_c)=0$.

Obviously, the BPS compacton solution which covers the whole fundamental domain $0\le f \le 1$ is completely different from the constant solution (\ref{const-sol}) even for a $V_{\mathbb{S}^2}$ very close to $V_B$. The phase transition from the constant solution to the BPS solution is, therefore, a first order transition in the case of compactons. Further, for compact BPS vortices the limit $V_\mathcal{M} \to V_B$ is not attainable. In this limit, the boundary of the compacton (with the topology of $\mathbb{S}^1$) would have to shrink to a point, which is incompatible with the nontrivial boundary condition for the Higgs field $u$ (it must wind $N$ times about the boundary for vortex number $N$).
 
\subsubsection{Case $G=(2/3)|1-|u|^2|/|u|^2$}
First of all, for $G=(2/3)|1-f^2|/f^2$ we observe that the Bradlow volume $V_B = 4\pi NG(0)=\infty$ is infinite. A related fact is that the constant solution (\ref{const-sol}) does not exist (i.e., has infinite energy).
Also, the corresponding potential does not obey the inequality (\ref{pot-ineq}). 
Unsurprisingly, thus, the solutions on $\mathbb{R}^2$ found in \cite{bazeia2016} behave in a way which is quite different from the case considered in the previous subsection. BPS solutions are still compact vortices, but the compacton radius now is $r_c = \sqrt{N(N+2)/3}$, leading to a compacton ``volume" $V_c = \pi r_c^2 = (\pi/3)N(N+2)$ which grows faster than linear (approximately quadratically) with the winding number. It is plausible to conjecture that there exists a relation between this unusual growth of the compacton volume, on the one hand, and the non-existence of a (finite) Bradlow volume, on the other hand, 
because any (finite) Bradlow volume proportional to $N$ would be surpassed by this compacton volume for sufficiently large $N$.  
On the other hand, compact BPS vortices will continue to exist at least on some compact manifolds without boundary provided their volume is sufficiently large. 
Indeed, from any compact vortex solution on $\mathbb{R}^2$ we immediately get a solution on a sufficiently large disc $\mathbb{D} \subset \mathbb{R}^2$ with the flat metric. From there we get a solution on a compact manifold with the topology of a two-sphere by just gluing this disc to a second disc along the boundaries of the two discs. The first disc may even be curved in the region outside the compacton. The gauge potential cannot be well-defined on the whole second disc (it is a large gauge transformation there), but this problem may be remedied by acting with the inverse of this large gauge transformation on $A_\mu$ and on $u$. It is well-known that on a compact manifold without boundary, in any case, one gauge potential cannot be defined globally and, instead, several potentials have to be chosen on several patches, related by large gauge transformations in the overlap regions (it is this fact which gives rise to the quantisation of the magnetic flux in the first place).
Solutions on more complicated manifolds may be constructed by gluing the first disc to more complicated surfaces with $\mathbb{S}^1$ boundaries instead of the second disc.
A different question is whether solutions exist on manifolds where the metric deviates from the flat one in the region of the compacton.
This problem is, however, beyond the scope of the present paper.

We conclude that, in this and related cases, a Bradlow bound as a useful criterion for the existence of BPS vortex solutions does not seem to exist
(the formal Bradlow volume is infinite, but BPS solutions on manifolds with finite volumes still exist). We could, of course, {\em define} the compacton volume for each $N$ as the Bradlow volume, but this is not a useful definition because {\it i)} it is not given a priori, i.e., we have to find the compact BPS solution and its volume to get the Bradlow volume and, {\em ii)} there is no guarantee that the Bradlow volume will be independent of the space manifold $\mathcal{M}$ (BPS solutions on different $\mathcal{M}$ might have different volumes). 

\section{Perfect fluid field theories}
Another class of models for which we will investigate the relation between the (geometric/thermodynamic) volume and the Bradlow volume is a family of models describing perfect fluids. This means that the corresponding action leads to an energy-momentum tensor of the perfect fluid type
\be
T^{\mu \nu}=(p+\rho) u^\mu u^\nu - p g^{\mu \nu}
\ee
where $p$ is the pressure, $\rho$ the proper energy density and $u^\mu$ the four-velocity. 
\subsection{Scalar field in 1+1 dimensions}
The simplest example of a perfect fluid type action is provided by a real scalar field in $(1+1)$ dimensions 
\be
\mathcal{L}=\frac{1}{2} \partial_\mu \phi \partial^\mu \phi - m^2 U(\phi) .
\label{scalar}
\ee
Then the energy-momentum tensor reads
\be
T_{\mu \nu}=\partial_\mu \phi \partial_\nu \phi -g_{\mu \nu} \mathcal{L}=u_\mu u_\nu (\mathcal{E} +\mathcal{P}) -g_{\mu \nu} \mathcal{P}
\ee
where the proper energy density and pressure density are
\be
\mathcal{E}=-\frac{1}{2} \partial_\mu \phi \partial^\mu \phi + m^2 U(\phi)
\ee
\be
\mathcal{P}=-\frac{1}{2} \partial_\mu \phi \partial^\mu \phi - m^2 U(\phi)
\ee
while the ``two-velocity" $u_\mu=(-\partial_\nu \phi \partial^\nu \phi)^{-1/2} \epsilon_{\mu \rho} \partial^\rho \phi$. For static configurations, the proper energy density and pressure density are simply components of the energy-momentum tensor
\be
\mathcal{E}=T_{00}=\frac{1}{2}\phi'^2 + m^2 U(\phi)
\ee
\be
\mathcal{P}=T_{11}= \frac{1}{2} \phi'^2 - m^2 U(\phi) .
\ee
Furthermore, due to the conservation of the energy-momentum tensor the pressure density has to take a constant value $\mathcal{P}=P$, which is simply an integration constant when we integrate the full second order static field equation once.
\\
Now we assume that the potential has at least two vacua at $\phi_+$ and $\phi_-$ which leads to the emergence of kinks (topological solitons) as configurations interpolating between them. Here the topological current is
\be
j_\mu =-\frac{1}{\phi_+-\phi_-} \epsilon_{\mu \nu} \partial^\nu \phi .
\ee
The energy of a kink is bounded from below by a topological bound 
\be
E=\frac{1}{2}\int_{-\infty}^{\infty} dx \left( \phi'^2+2m^2U\right) \geq \left| m \int dx \sqrt{2U}\phi'\right| =m
\left| \int_{\phi_-}^{\phi_+} d\phi \sqrt{2U}\right|
\ee
where the last integral is over the ``fundamental domain" of target space (the region $\phi_- \le \phi \le \phi_+$ where the soliton takes its values). The bound is saturated by solutions of the zero-pressure equation called the BPS equation
\be
\frac{1}{2} \phi'^2 - m^2 U(\phi)=0.
\ee
Such BPS kinks are global energy minima in the respective topological sectors. However, kink solutions exist also for any positive value of the pressure. Their energy is again given by a target space (solution independent) formula
\be
E(P)=\frac{1}{\sqrt{2}} \int_{\phi_-}^{\phi_+} d\phi \frac{2m^2U+P}{\sqrt{2m^2U+P}} .
\ee
Now, the vacua are approached at a {\it finite} distance leading to a finite geometric volume 
\be
V(P)=\frac{1}{\sqrt{2}} \int_{\phi_-}^{\phi_+} d\phi \frac{1}{\sqrt{2m^2U+P}} .
\ee
As we know this volume is the thermodynamic volume, because
\be
P=-\frac{dE}{dV}
\ee
At the equilibrium $(P=0)$  
\be
\phi' = \pm m\sqrt{2U} \;\; \Rightarrow \;\; \frac{d\phi}{m\sqrt{2U}} =dx
\ee
where the last formula makes sense only where the field is outside of the vacuum. Then the geometric volume
\be
 V=\frac{1}{\sqrt{2}} \int_{\phi_-}^{\phi_+}  \frac{d\phi}{\sqrt{2m^2U}}
\ee
can be infinite (usually infinitely extended solitons) or finite (compactons), which is completely governed by the potential or, more precisely, by its behavior close to the vacua. 

The volume of a scalar soliton in (1+1) dimensional models is base space independent. In fact, as the expression of the volume is a target space integral, which obviously is not sensitive to a particular form of the solution, the result is exactly the same, whether the base space is $\mathbb{R}$ or $\mathbb{S}^1$ with radius $R$. Of course, the only limitation is that the solution does exist i.e., the Bogomolny equation has a pertinent soliton on a finite volume space. This is equivalent to the appearance of the Bradlow type bound. A BPS soliton exists if its geometrical volume is smaller than the volume of the base space manifold. Quite interestingly, the upper geometric volume is identical to the Bradlow volume (minimal volume in which BPS kinks exist)
\be
V_B=\frac{1}{\sqrt{2}} \int_{\phi_-}^{\phi_+}  \frac{d\phi}{\sqrt{2m^2U}} .
\ee
Indeed, for compactons one can always put them in a bigger volume $V' > V_B$ as a superposition of non-overlapping smaller charge units. This is a rather simple thermodynamical system of a gas of non-interacting compactons. Hence, the bound is now
\be
V_B \leq V' \leq V_\mathcal{M}
\ee
If we are below the bound no BPS solitons exist. Topological solitons become now non-BPS objects solving the non-zero pressure equation. 

\vspace*{0.2cm}

Let us underline that the Bradlow volume is finite if and only if the kink is a compacton (for example in $\mathbb{R}$). This is uniquely determined by the potential or, more precisely, by the approach to the vacuum. If close to vacuum $\phi = \phi_a+\delta \phi$ and $U \propto (\delta \phi)^c $, then compactons exist if $c<2$.  On the other hand, the Bradlow volume is infinite if and only if kinks are usually infinitely extended solitons in $\mathbb{R}$. Note that in the Abelian Higgs model at the critical coupling the geometric volume of a BPS vortex on $\mathbb{R}^2$ is infinity although the Bradlow volume is finite. 

\vspace*{0.2cm}

In the next sections we prove that the features outlined above are genuine features of any perfect fluid scalar field theory in any dimension. 
\subsection{Higher-dimensional generalisation}
There is a generalisation of the ($1+1$) dimensional single scalar theory to any dimension which preserves the perfect fluid property. It relies on the observation that the Lagrangian (\ref{scalar}) can be written as
\be
\mathcal{L}=-\frac{(\phi_+-\phi_-)^2}{2} j_\mu j^\mu - m^2 U(\phi) 
\ee
where $j_\mu$ is the topological current. Then, in $(d+1)$ dimensions the models we want to consider are \cite{BPSgen}
\be
\mathcal{L}=-\frac{1}{2}j_\mu j^\mu - m^2 U
\ee
where now we assume a target space spanned by d scalars $\phi=(\phi^a)$, $a=1...d$ and 
\be
j_\mu = \frac{1}{d!}\epsilon^{a_1...a_d} \epsilon_{\mu \mu_1...\mu_d} K(\phi^a)  \partial^{\mu_1}\phi^{a_1}...\partial^{\mu_d} \phi^{a_d}
\ee
where $K$ is related to a volume form on the target space $\mathcal{N}$
\be
d\Omega^{d} = K(\phi^a) d\phi^1\wedge...\wedge d\phi^d
\ee
For the static case, the total energy is again bounded from below by a Bogomolny type bound
\be
E= \frac{1}{2} \int d^dx \left( j_0^2+2m^2U \right) \geq \left| \int d^d x j_0 \sqrt{2mU} \right| = k \int_{\mathcal{N}'} d\Omega^d \sqrt{2mU}
\ee
where $k$ is a winding number of the map and $\mathcal{N}'$ is the fundamental domain of the soliton which can but does not have to coincide with the whole target space. In fact, in the examples discussed below we assume $\mathcal{N}=\mathbb{C}$ while $\mathcal{N}'=\mathbb{D}$ is a disc. The bound is saturated by solutions of a Bogomolny equation
\be
j_0^2-2m^2 U=0.
\ee
Furthermore, the models lead to a perfect fluid energy-momentum tensor which, in the static case, has the non-zero components 
\be
T_{00} =\mathcal{E}=\frac{1}{2}j_0^2 + m^2U
\ee
\be
T_{ii} =\mathcal{P}=\frac{1}{2}j_0^2 - m^2U
\ee
where again due to the energy-momentum tensor conservation law $\mathcal{P}=P=const.$ This gives rise to non-zero pressure configurations which solve the equation
\be
\frac{1}{2}j_0^2 =P+ m^2U
\ee
where the vacuum value must be approached at a finite distance (volume)
\be
V=k \int_{\mathcal{N}'} d\Omega^d \frac{1}{\sqrt{2mU+P}}.
\ee
Finally we can also define the Bradlow volume
\be
V_B=k \int_{\mathcal{N}'} d\Omega^d \frac{1}{\sqrt{2mU}} .
\ee
In the subsequent parts of the paper we will provide some examples of such models and discuss them from the Bradlow volume and Bradlow bound perspective. 
\subsection{Global vortices with the SDiff Symmetry}
\subsubsection{Model and Bogomolny equation}
Perhaps the best and simplest example of a solitonic model with a perfect fluid energy-momentum tensor in (2+1) dimensional space-time is provided by the following model  \cite{BPSvortex}
\begin{equation}
\mathcal{L}_{BPS \; vortex}=\mathcal{L}_4+\mathcal{L}_0 \label{bps vortex}
\end{equation}
which is closely related to the so-called BPS baby Skyrme model \cite{bps-baby}-\cite{bps-baby Z}.
It consists of two parts:  a fourth derivative term 
\begin{equation}
\mathcal{L}_4=-(u_\mu \bar{u}^\mu)^2 + u_\mu^2\bar{u}_\nu^2,
\end{equation}
and a non-derivative part, i.e., a potential
\begin{equation}
\mathcal{L}_0= -U(u,\bar{u})
\end{equation}
which has a vacuum at $|u|=1$. 
Without loss of generality we set all constants equal to 1. It can be proven that there is a topological bound for the static energy for any potential $U$ 
\begin{eqnarray}
E_{BPS\; vortex} &=& \int_\mathcal{M} \sqrt{g}  d^2x \left( [(\nabla u \nabla \bar{u})^2  - (\nabla \bar{u})^2 (\nabla u)^2] +U \right)
\nonumber \\ &=&
\int_\mathcal{M} \sqrt{g} d^2x \left( \frac{i}{\sqrt{g}}\epsilon_{ij} \partial_i u \partial_j \bar{u}  \pm  \sqrt{U} \right)^2 \mp  2 i  \int_\mathcal{M} d^2x \epsilon_{ij} \partial_i u \partial_j \bar{u} \sqrt{U} \\
&\geq &\vert 2i  \int_\mathcal{M} d^2x  \epsilon_{ij} \partial_i u \partial_j \bar{u} \sqrt{U} \vert \\
&=&4\pi \vert N\vert  \langle \sqrt{U}\rangle_{\mathbb{D}}
\end{eqnarray}
as the topological charge is
\be
N=\frac{i}{2\pi} \int d^2x \epsilon_{ij} \partial_i u \partial_j \bar{u} \equiv \int d^2 x q.
\ee
where $q$ is the charge density. The disc $\mathbb{D} = \{u \in \mathbb{C} \; \big| \; |u|   \le 1 \}$, a subspace of the full $\mathbb{C}$ target space, is the fundamental domain of the vortex solution. Moreover,
\be
 \langle \sqrt{U}\rangle_{\mathbb{D}} = \int \frac{idu \wedge d\bar{u}}{2\pi} \sqrt{U} =\frac{1}{\pi}\int fdf d\phi \sqrt{U} 
\ee
is the average value of $\sqrt{U}$ on $\mathbb{D}$. Here, $f, \phi$ are polar coordinates on $\mathbb{D}$ i.e., $u=f{\rm e}^{i\phi}$.
The bound is saturated if and only if the following Bogomolny equation is satisfied
\begin{equation}
 \frac{i}{\sqrt{g}}\epsilon_{ij} \partial_i u \partial_j \bar{u}=\pm  \sqrt{U} .\label{Bog eq}
\end{equation}
It can be shown that the saturating solutions do fulfil the second order Euler-Lagrange equations. So, the model possesses a BPS sector, which is not empty if the Bogomolny equation has at least one solution. Note, that identically as in the 1-dimensional case the Bogomolny equation is defined for any potential. This is a fundamental difference compared to the Abelian Higgs model. 
\\
In the subsequent analysis we will consider a family of potentials 
\be
U=\frac{1}{4} \left( 1-u\bar{u} \right)^{2\alpha},
\ee
where the parameter $\alpha \geq 1/2$. 
\subsubsection{Example - flat space}
Before we discuss the Bradlow bound and the Bradlow volume in a general set-up it is instructive to solve the Bogomolny equation exactly in the case where $U$ depends only on $|u|$. 
In flat space one may perform it assuming the usual axially symmetric ansatz 
\begin{equation} 
u(r,\varphi) = f(r) {\rm e}^{iN\varphi}
\label{ans}
\end{equation}
where $N \in \mathbb{Z}$ is the topological charge (winding number) and $f$ is a profile function. Here we used polar coordinates. Note that further (but not necessarily all) solutions can be obtained if {\it SDiff} transformations are applied to the axially symmetric solutions. 
Now, we have
\be
\frac{2N}{r} ff_r=\frac{1}{2} (1-f^2)^\alpha .
\ee
A topologically nontrivial solution requires $f(r=0)=0$ and $f(r=R)=1$ where, as we will see, the soliton boundary $R$ can be
infinite (as for usual solitons) or finite (compactons). Then, depending on the value of the parameter $\alpha$ we find the following
types of solutions. For $\alpha \in \left(\frac{1}{2},1\right)$ we have compactons
\begin{equation} \label{sdiff-vort-R2}
1-f^2= \left\{
\begin{array}{cc}
\left( 1-\frac{r^2}{R^2}\right)^{\frac{1}{1-\alpha}} & r \leq R
\\
 & \\
 0 & r \geq R,
\end{array}
\right. \;\;\;\;\;\;\; R^2=\frac{4N}{1-\alpha} .
\end{equation}
Here $R$ is the compacton radius.
For $\alpha=1$ we find (more than) exponentially localized solitons
\be
1-f^2={\rm e}^{-\frac{r^2}{4N}} .
\ee
Finally for $\alpha >1$ we obtain power-like localized solitons  
\begin{equation}
1-f^2=\left(\frac{R^2}{r^2+R^2} \right)^{\frac{1}{\alpha-1}}, \;\;\;\;\;\; R^2=\frac{4N}{\alpha-1} .
\end{equation}
Here $R$ is just an integration constant (not a compacton radius as the solitons extend to infinity). The conclusion is that in the flat space the BPS sector is not empty. The Bogomolny equation possesses solutions for any value of $\alpha$ i.e., for any possible power-type approach to the vacuum. 
\subsubsection{Example - finite volume space $\mathcal{M}=\mathbb{S}^2 \backslash \{ (0,0,-1)\}$}
As we found compact vortex solutions for some potentials (some values of $\alpha$) on $\mathbb{R}^2$, one might naively expect similar compact solutions on manifolds with a finite but sufficiently big volume ($V_\mathcal{M} > V_B$).  
There is, however, a topological obstruction to the existence of these solutions which requires the manifold to be noncompact, or have a boundary. The reason is that, for vortices, the vacuum is at $|u|=1$ which implies that, outside the domain of the compacton, $u=e^{i\phi (x^j)}$. 
Further, for vortices with vortex number $N$, the phase $\phi (x^j)$ must change by $2\pi N$ along closed paths which enclose the domain of the compact vortex. But on a compact manifold without boundary this is impossible. For a manifold with the topology of a two-sphere, e.g., one may shrink such a curve to a point in the hemisphere opposite to the compacton, which is obviously incompatible with the nontrivial winding. The way out is a Higgs field $u$ which has a singularity at some point in the ``vacuum hemisphere" which impedes the shrinking. This point must then be removed from the manifold, giving it the topology of an open disc.  Alternatively,
one can remove an open disc from $S^2$, leaving a compact manifold with boundary, diffeomorphic to a closed disc.
For gauged {\em SDiff} vortices, the problem with the nontrivial phase $u=e^{i\phi (x^j)}$ in the vacuum region continues to exist, because it will turn out that the magnetic flux of the corresponding BPS solutions is not quantised. We remark that this problem does not exist for BPS baby skyrmions ({\em SDiff} baby skyrmions), because there the unique vacuum value is $u=0$, which may be extended to the whole vacuum region (region outside the compacton domain) on arbitrary manifolds. So our considerations below, with some small modifications, apply to that case even for compact manifolds without boundary.

So, let us consider, for simplicity, the two-dimensional sphere with the south pole removed, $\mathcal{M}=\mathbb{S}^2\backslash \{ (0,0,-1)\}$.
The metric still is
\be
ds^2_{\mathbb{S}^2} = R^2_{\mathbb{S}^2}(d\theta^2 + \sin^2 \theta d\phi^2) ,
\ee
but now $0 \le \theta < \pi$. 
Using the ansatz $u(\theta,\phi) = f(\theta) {\rm e}^{iN\varphi} $ we find the Bogomolny equation in the form
\be
\frac{2N}{R^2_{\mathbb{S}^2}\sin \theta} ff_\theta=\frac{1}{2} (1-f^2)^\alpha .
\ee
The boundary conditions are $f(\theta=0)=0$ and $f(\theta=\theta_c)=1$, where now $\theta_c$ must be smaller than $\pi$. This is a nontrivial restriction which puts some bounds on the existence of BPS vortices for our equation. 

We start with $\alpha \in \left(\frac{1}{2},1\right)$. Then, 
\begin{equation} \label{sdiff-vort-S2}
1-f^2= \left\{
\begin{array}{cc}
\left(\frac{\cos \theta - \cos \theta_c}{1-\cos \theta_c}\right)^{\frac{1}{1-\alpha}} & \theta \leq \theta_c
\\
 & \\
 0 & \theta \geq \theta_c,
\end{array}
\right. \;\;\;\;\;\;\; \cos \theta_c =1-\frac{2N}{R^2_{\mathbb{S}^2}(1-\alpha)}.
\end{equation}
This formal solution makes sense only if $1\geq \cos \theta_c > -1$. The first inequality is always satisfied. However, the second one provides a restriction (for a given potential $\alpha <1$) for the topological charge and radius of the two sphere i.e., the area (volume) of the compact manifold  
\be
1-\frac{2N}{R^2_{\mathbb{S}^2}(1-\alpha)}> -1 \;\; \Rightarrow \;\; V_{\mathcal{M}} > \frac{4\pi}{1-\alpha}  \vert N\vert .
\ee
Only for a sphere with the volume larger than $\frac{4\pi}{1-\alpha}  \vert N\vert $ a BPS solution with topological charge $N$ exists.  
This is exactly the Bradlow bound for the {\em SDiff} BPS global vortices. 

For $\alpha=1$ one can easily solve the BPS equation 
\be
1-f^2=C{\rm e}^{-\frac{R^2_{\mathbb{S}^2}}{2N} \cos \theta} .
\ee
The problem is that there is no $C$ for which the boundary conditions, $f(\theta=0)=0$ and $f(\theta=\theta_c < \pi)=1$, would be satisfied. Therefore the corresponding topologically nontrivial BPS vortex cannot exist. The same happens for any $\alpha >1$. 
\\
We conclude that for $\alpha \geq 1$ (which for the flat space corresponds to infinitely extended vortices) the BPS sector on 
$\mathbb{S}^2 \backslash \{ (0,0,-1)\}$ is empty. 
\subsubsection{The Bradlow bound}
It is not difficult to derive the pertinent Bradlow bound (and corresponding Bradlow volume) for general potential $U$ and any manifold $\mathcal{M}$ with the right topology. Let us assume for the moment that BPS solutions on a given manifold exist. Then we divide the Bogomolny equation (\ref{Bog eq}) by $\sqrt{U}$, which makes sense only outside the vacuum  
\begin{equation}\label{ranhas1}
2\pi\frac{1}{\sqrt{U}} \frac{ i}{2\pi} \frac{1}{\sqrt{g}}\epsilon_{ij} \partial_i u \partial_j \bar{u} =1. 
\end{equation}
Now we integrate it over the base space manifold (remembering that the equation is valid only inside the geometric volume of the soliton)
\begin{equation}
2\pi \int_\mathcal{M}\frac{1}{\sqrt{U}} \frac{ i}{2\pi} \epsilon_{ij} \partial_i u \partial_j \bar{u} d^2 x=\int_\mathcal{M} \sqrt{g} d^2 x
\end{equation}
Hence, 
\begin{equation} 
2 N \int  \frac{1}{\sqrt{U}}  \mbox{vol}_{\mathbb{D}}=V
\end{equation}
or 
\be
2\pi N \left\langle \frac{1}{\sqrt{U}} \right\rangle_{\mathbb{D}}=V
\ee
where $V$ denotes the geometric volume of the BPS soliton. In fact, the left hand side of this expression is the Bradlow volume
\be
V_B=2\pi N \left\langle \frac{1}{\sqrt{U}} \right\rangle_{\mathbb{D}} .
\ee
Observe that the target space average of $\mathcal{U}^{-1/2}$ diverges if the potential gives infinitely extended vortices in flat space, while it is finite in the case of compactons. Furthermore, for compactons it is trivial to construct a solution with the same topological charge but in a bigger volume $V'$. One has to put for example $N$ charge one compact vortices such that there is the vacuum between them. Then, such a configuration still is a BPS solution. Obviously, the volume cannot be larger than the volume of the manifold, 
\be
V_B \leq V' \leq V_{\mathcal{M}},
\ee
which is precisely the Bradlow bound with the exactly computed Bradlow volume. The energy of all such non-overlapping BPS vortices is constant which can be interpreted as a gas of noninteracting solitons with $P=0$. In contrast to the Abelian Higgs vortices, such BPS configurations with $V>V_B$ are trivial to construct. 

The meaning of this law is as for the Abelian Higgs vortices and for kinks in 1+1 dimensions. An {\it SDiff} BPS vortex may exist if its Bradlow volume is smaller that the volume of the base space manifold, i.e., the manifold $\mathcal{M}$ is large enough to host the BPS vortex. If the Bradlow volume (minimal volume of charge $N$ soliton) $V_B$ is bigger that $V_\mathcal{M}$ then one can still have vortex solutions on $\mathcal{M}$. However, such vortices will no longer be solutions of the Bogomolny equation (not BPS vortices) but of the full Euler-Lagrange (EL) equations. For the perfect fluid model, the following generalised first order equation 
\begin{equation} \label{sdiff-p-eq}
2\pi \frac{ i}{2\pi} \frac{1}{\sqrt{g}} \epsilon_{ij} \partial_i u \partial_j \bar{u} =\sqrt{U+P} 
\end{equation}
is a first integral of the static EL equations, where the pressure $P$ is an integration constant.
Therefore, 
\be 
2\pi N \left\langle \frac{1}{\sqrt{U+P}} \right\rangle_{\mathbb{D}} = V.
\ee
For example, for potentials for which $V_B$ is infinite (usually infinitely extended solitons) the {\it SDiff} BPS vortices can be only constructed on an infinite volume manifold. If we force them on a finite manifold then they will not be BPS solitons following the Bogomolny equation but vortex solutions of the full E-L system, which here means the same as solutions of the generalized Bogomolny equation with non-zero pressure.

So far, the considerations of this section have {\em assumed} that solutions to Eq. (\ref{sdiff-p-eq}) exist (BPS solutions for $P=0$ and non-BPS ones for $P>0$). 
We now want to show that such solutions do exist, on sufficiently large domains $\MM$, under mild topological restrictions. In order to include the case $P=0$, we assume $U$ is chosen so that $V_B<\infty$, implying that BPS solutions (if they exist) are compactons. We also restrict to the case $N=1$.
Note that (\ref{ranhas1}) is just a special case of (\ref{sdiff-p-eq}), and that the latter has a natural geometric
interpretation \cite{bps-baby Sp1}. Given a pressure $P\geq 0$ solution of the EL equations $u:\MM\ra\C$, let $\MM_{vac}=\{x\in\MM\: :\: |u(x)|=1\}$ and $\MM'=\MM\less\MM_{vac}$. Then (\ref{sdiff-p-eq}) implies that $u$ defines a volume preserving map $\MM'\ra\DD'$, where $\DD'$ is the unit disc equipped with the deformed volume form
$\Omega=\frac{i}{2}\frac{dz\wedge d\bar{z}}{\sqrt{U(z,\bar{z})+P}}$. Conversely, given a volume preserving map
$u':\MM'\ra\DD'$ from $\MM'\subseteq\MM$ of volume $V$, this defines a pressure $P\geq 0$
vortex solution {\em provided it can be continuously extended} to $\MM$ so that $|u|=1$ on
$\MM_{vac}:=\MM\less\MM'$.

So, let $\MM$ be an oriented Riemannian two-manifold (possibly with boundary) of volume greater than $V$. Then it certainly contains a submanifold $\MM'$, diffeomorphic to an open disc, with volume $V$. Furthermore, there exists a volume preserving diffeomorphism $u':\MM'\ra\DD'$ \cite{moser}. Since $u'$ has no critical points, it extends 
uniquely to a map $\ol{u'}$ from the closure $\ol{\MM'}$ to the closure $\ol{\DD}$ which maps the boundary $\S^1=\cd{\MM'}$ homeomorphically, and with winding one, to the unit circle $\S^1=\cd\ol{\DD}$. As above, let $\MM_{vac}=\MM\less
\MM'$. Assume that there exists a continuous map $f:\MM_{vac}\ra\cd\MM'$ such that for all $x\in\cd\MM'$, $f(x)=x$. Then the volume preserving map $u':\MM'\ra\DD'$ has a continuous
extension to $\MM$ given by $u(x)=(\ol{u'}\circ f)(x)$ for all $x\notin\MM'$ which, by definition, has $|u(x)|=1$ for all $x\notin\MM'$. This, then, is a charge $1$ pressure $P\geq 0$ vortex solution on $\MM$. 

\begin{figure}
\begin{center}
\includegraphics[scale=0.5]{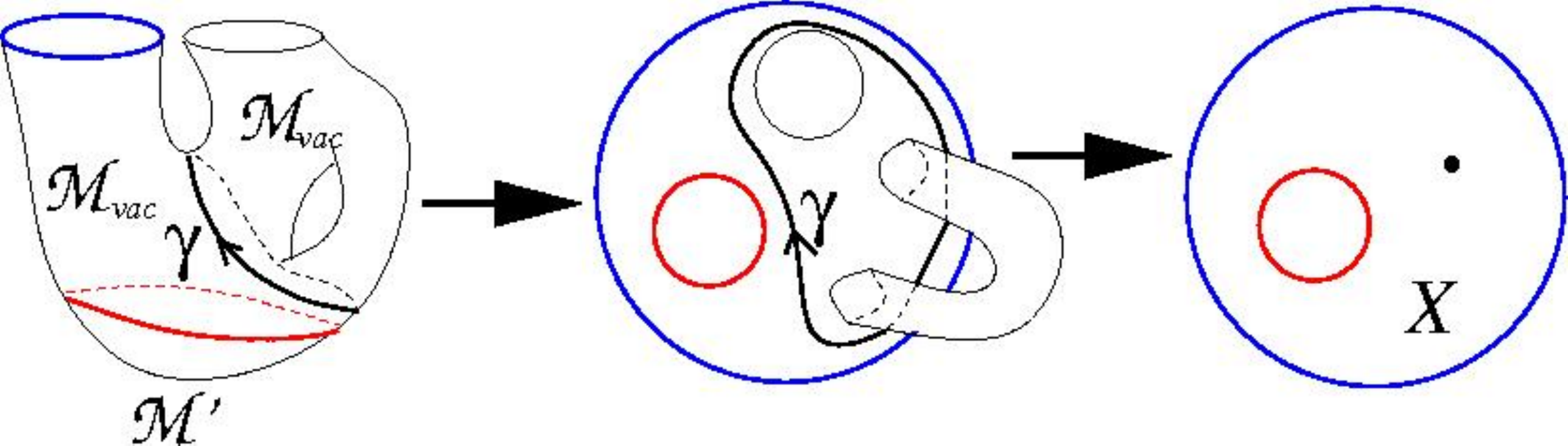}
\end{center}
\caption{ Extending a volume preserving diffeomorphism $\MM'\ra\DD'$ to a global map $\MM\ra\ol{\DD}$ via a collapse and retraction construction.}
\label{fig:ranhas}
\end{figure}

Under what circumstances does a map $f:\MM_{vac}\ra\MM_{vac}$ with the required properties exist? Recall 
\cite{hatcher} that a
{\em retraction} of a topological space $X$ to a subspace $A$ is a continuous map $f:X\ra A\subset X$ with 
$f(x)=x$ for all $x\in A$ {\em which is homotopic to the identity map $\id:X\ra X$}. Hence, a retraction from $X=\MM_{vac}$ to the subspace $A=\cd\MM'\subset\MM_{vac}$ certainly has the required property to define an extension of $u'$, and so an extension certainly exists if $\MM_{vac}$ retracts to the circle
$\cd\MM'$. This happens if $\MM\cong\R^2$, since then $\MM_{vac}\cong \R^2\less\DD \cong \ol{\DD}\less\{0\}$
which retracts (in fact, strongly deformation retracts) to its boundary. The only other case in which a retraction exists is $\MM\cong\ol{\DD}$, a closed disk, for which $\MM_{vac}$ is homeomorphic to a closed annulus, which, again, strongly deformation retracts to its inner boundary circle. Note, however, that requiring $f:\MM_{vac}\ra \MM_{vac}$ to be  a retraction is really overkill: there's no reason to require $f$ to be homotopic to the identity map. Consider the case where $\MM$ is compact with boundary. Then $\MM$ is diffeomorphic to $S^2$ with at least
one open disk removed, and some number of handles attached. Remove from $\MM$ the open disk $\MM'$ to obtain
$\MM_{vac}$, which is a topologically a sphere with at least two open disks removed, and some number of handles. From $\MM_{vac}$ we can construct a topological space $X$ homeomorphic to a cylinder (equivalently, a closed annulus) as follows: draw a simple closed curve $\gamma$ in the interior of $\MM_{vac}$ which encloses both $\cd\MM'$ and
one of the boundary circles of $\MM$.  Now define the equivalence relation $\sim$ on $\MM_{vac}$ which collapses all points outside $\gamma$ to a single point. Then $X=\MM_{vac}/\sim$ is homeomorphic to a closed annulus, and the canonical map $\pi:\MM_{vac}\ra X$, $\pi(x)=[x]$ is tautologically continuous. As previously argued, a retraction
$f_X$ from 
$X$ to its inner boundary circle exists. But then $f=f_X\circ\pi$ is a continuous map $\MM_{vac}\ra\cd\MM'$ which
restricts to the identity on $\cd\MM'$, and hence, defines an extension of $u'$. The construction is illustrated in figure \ref{fig:ranhas}. Note that the extension is not unique. In fact, if $\MM$ has more than one boundary component, the extension is not even unique up to homotopy, since the curve $\gamma$ can encircle any of the boundary circles. Note also that we can equally well start with $\MM$ diffeomorphic to a sphere with at least one {\em point} (rather than open disc) removed, and any number of handles attached, and take $\gamma$ to enclose exactly one puncture point and $\MM'$. We then obtain $X$
homeomorphic to a punctured disc which, again, retracts to its boundary. We could also remove some points and some open disks. We see, then, that charge 1, pressure $P\geq 0$ vortices exists on manifolds $\MM$ of sufficiently large volume for essentially any imaginable topology, except compact manifolds without boundary.

From the point of view of the Bradlow bound, there are some similarities between {\it SDiff} BPS vortices and Abelian Higgs vortices. There is always a Bradlow volume i.e., minimal volume for which a BPS solution may exist. If we further decrease the volume of the vortex (by squeezing the solution or by reducing the volume of the manifold) no BPS configurations can exist. Instead,  non-BPS vortices may appear, which leads to non-zero pressure. A more careful analysis, however, reveals some interesting differences.
Firstly, global {\em SDiff} vortices cannot exist on compact manifolds without boundary, because the nontrivial vacuum field $u=e^{i\phi}$ cannot be extended to the whole vacuum region on such manifolds. As said, this problem does not exist for 
{\em SDiff} baby skyrmions and {\em SDiff} skyrmions. Secondly,
similar to the case of 1+1 scalar field theories, the geometric volume $V$ of a {\it SDiff} BPS vortex is always equal to the Bradlow volume. Therefore, only compactons (whose existence is again completely determined by the potential) lead to a finite Bradlow volume. 
Thirdly, the (local) existence of a compact vortex on a finite (non-vacuum) domain on any sufficiently large manifold is guaranteed by the fact that the corresponding first-order equation is independent of the metric.
All these findings, obviously, continue to hold for further versions of {\em SDiff} BPS models like, e.g., the BPS baby Skyrme model or the BPS Skyrme model  \cite{BPSskyrme} in 3+1 dimensions. It follows that solutions for these further models exist on arbitrary space manifolds, e.g., on compact manifolds without boundary.
\section{Abelian vortices with the SDiff Symmetry}
\subsection{Model and Bogomolny equations}
Finally, we extend our analysis for a gauged version of perfect fluid theory which we call {\it SDiff} Abelian vortex model \cite{BPSvortex}
\begin{equation}
\mathcal{L}_{gauged \;BPS \;vortex}=- \frac{1}{2}[ (D_\mu u \overline{D^\mu u})^2- (D_\mu u)^2 (\overline{ D_\nu u})^2] - U(u\bar{u}) - \frac{1}{4e^2} F_{\mu \nu}^2,
\end{equation}
where the usual derivatives are replaced by covariant derivatives given by
\begin{equation}
D_\mu u = u_\mu +i A_\mu u .
\end{equation}
We now keep the constant $e$ (the dimensionless electric charge). Here we repeat the computation of the topological bound and the corresponding Bogomolny equations saturating the bound \cite{BPSvortex} (which is a version of the computations originally done for the gauged BPS baby Skyrme model \cite{gauged bps}).

The static energy of the SDiff Abelian model is given by
\begin{equation}
  E = \frac{1}{2}\int \sqrt{g} d^2x \left( \frac{Q^2}{g} + 2U + \frac{1}{e^2} \frac{B^2}{g} \right),
\end{equation}
where the covariant ``topological density" $Q$ takes the form
\begin{eqnarray}
\frac{Q}{\sqrt{g}} &\equiv & \frac{i}{\sqrt{g}} \epsilon_{ij} D_i u \overline{D_j  u} = \frac{i}{\sqrt{g}} \epsilon_{ij} u_i \bar u_j - \frac{1}{\sqrt{g}}\epsilon_{ij} A_i \partial_j |u|^2 
\equiv 2\pi \frac{q}{\sqrt{g}} -  \frac{1}{\sqrt{g}} \epsilon_{ij} A_i \partial_j |u|^2 \\
&=&  \frac{1}{\sqrt{g}} \epsilon_{ij} \partial_i |u|^2  (\phi_j + A_j) = -\frac{1}{\sqrt{g}}\epsilon_{ij} \partial_i h (A_j + \phi_j) 
\end{eqnarray}
where 
\be
u=f{\rm e}^{i\phi}, \;\;\; h\equiv 1-f^2 .
\ee
Let us now consider a suitable non-negative expression
\begin{eqnarray}
0 &\le &  \left(\frac{Q}{\sqrt{g}} - w(h)\right)^2 + \frac{1}{e^2}\left(\frac{B}{\sqrt{g}}+b(h) \right)^2 \nonumber \\
&=&  \left(\frac{Q^2}{g} + w^2 \right) + \frac{1}{e^2}\left( \frac{B^2}{g} + b^2 \right) - 2\cdot 2\pi\frac{q}{\sqrt{g}} w \nonumber \\
&& +\frac{2}{\sqrt{g}}\epsilon_{ij} w(h) \partial_i h A_j + \frac{2}{e^2} b(h)\frac{1}{\sqrt{g}}\epsilon_{ij} \partial_i A_j 
\end{eqnarray}
where $w(h)$ and $b(h)$ are new functions of the target space variable $h$ that are still to be defined. The last two terms in the upper expression can be combined into a total derivative if we assume  that these functions obey the following relation
\begin{equation}
b(h) = e^2 W(h) , \quad W(h) \equiv \int_0^h dh' w(h') 
\end{equation}
\begin{equation}
 \Rightarrow \quad  
 \epsilon_{ij} w(h) \partial_i h A_j + \frac{1}{e^2} b(h)\epsilon_{ij} \partial_i A_j = \epsilon_{ij} \partial_i (WA_j).
\end{equation}
Of course, by construction $W(h)$ is zero at the vacuum value $h=0$ and therefore the total derivative does not contribute to the energy and may  be omitted. The remainder of the non-negative expression may indeed be written as the energy density minus the topological term $2  q W_h$ provided that the function $W$ obeys the first order nonlinear ODE (the ``superpotential equation")
\begin{equation} \label{superpot-eq}
 W_h^2 +   e^2 W^2 = 2U(h).
\end{equation}
Assuming that this is the case  we find for the energy the inequality
\begin{equation}
E=\frac{1}{2}\int \sqrt{g} d^2 x \left( \left(\frac{Q}{\sqrt{g}}  - W_h \right)^2 + \frac{1}{e^2} \left(\frac{B}{\sqrt{g}}  - e^2 W\right)^2 \right) +    2\pi \left\vert \int d^2 x  qW_h \right\vert
\ge    2\pi \left\vert \int d^2 x  q W_h \right\vert
\end{equation}
Hence
\begin{equation}
E
\ge  2\pi \vert N \vert  \left\langle W_h \right\rangle_{\mathbb{D}}
\end{equation}
The bound is saturated if the following Bogomolny equations are satisfied
\begin{equation}
 \frac{Q}{\sqrt{g}} =W_h \; ,\quad \frac{B}{\sqrt{g}} = -e^2   W \label{g bog}
\end{equation}
Again, one can prove that solutions of the Bogomolny equations solve the full second order field equations. 
\subsection{Bradlow bound for the gauged BPS vortices}

For the derivation of the volume of a soliton in the {\em SDiff} BPS type models (non-gauge case) we 
simply integrated the BPS equation over the manifold $\mathcal{M}$. So, one would like to do this also in the gauged 
version. Unfortunately, as the magnetic and matter (vortex) field are nontrivially mixed in the Bogomolny equations, we were not able to perform this computation in an ansatz-independent manner. However, since the model in the static version has {\it SDiff} invariance, a result obtained in a certain ansatz may be easily generalised to any {\it SDiff} related configuration, which at the end renders our result ansatz (and coordinate) independent. As in the case of global vortices, we will first consider the case of compact solutions on $\mathbb{R}^2$, and then invoke the metric independence of the BPS equations (\ref{g bog}) to argue that the same solutions continue to exist (at least locally, i.e., on a compact non-vacuum domain) on arbitrary (sufficiently large) manifolds for an appropriate choice of coordinates (leading to the same volume element $g$).

We consider flat space and assume the ansatz 
\be
u(r,\varphi) = f(r) {\rm e}^{iN\varphi}, \;\;\;\; A_0=A_r=0, \; A_\varphi= Na(r) .
\ee 
The Bogomolny equations take the following form
\begin{equation} \label{BPS1-spher}
Nh_y (1+a) = -W_h
\end{equation}
\begin{equation} \label{BPS2-spher}
Na_y = -e^2 W
\end{equation}
where, as before, $h=1-f^2$ and $y=r^2/2$. First we compute the magnetic flux, which can be expressed in terms of the asymptotic value of the gauge potential 
\be
\Phi = \int rdrd\varphi B = 2\pi N \int dy a_y = 2\pi N a(y_0) \equiv 2\pi N a_\infty .
\ee
Here $y_0$ is the position of the soliton boundary, which is finite for compactons and infinite for non-compact solitons. Dividing one Bogomolny equation by the other we get
\be
\frac{a_y}{1+a} = e^2 h_y \frac{W}{W_h}
\ee
Hence
\be
\partial_y \ln (1+a) = e^2  \partial_y F
\ee
where 
\be
F_h \equiv \frac{W}{W_h}  
\ee
The last expression can be formally integrated 
\be
F (h) = \int_0^h dh' \frac{W(h')}{W_h (h')}
\ee
which results in 
\be
\ln C(1+a) = e^2 F(h(y)) 
\ee
where $C$ is an integration constant. Here we assume that the first derivative of the superpotential does not possess zeros in the interval $0<h\le 1$. This guarantees that $F_h$ is finite in the same interval. At the vacuum $h=0$, where $W_h =0$, we assume that the potential behaves algebraically, i.e. $V \sim h^{2\alpha}$ for some $\alpha >0$, then $W_h \sim h^\alpha $, $W \sim h^{\alpha +1}$ near $h=0$ and $F_h$ is, in fact, zero at $h=0$. As a consequence, $F$ exists and is finite in the whole interval $h\in [0,1]$. For such (generic) potentials, it follows from the above result that 
\be
a(y) > -1, \;\;\; \mbox{for all} \;\; y 
\ee
and that the limit $a\to -1$ may be reached only in the limit $e \to \infty$. The integration constant may be determined from the boundary conditions $h(y=0)=1$, $a(y=0)=0$,
\be
F(1)=\frac{1}{e^2} \ln C \quad \Rightarrow \quad C={\rm e}^{e^2 F(1)}
\ee
which, together with $h(y_0)=0$ and $F(h=0)=0$ leads to the asymptotic expression
\be
a_\infty =-1+{\rm e}^{-e^2 F(1)}
\ee
which may be inserted into the expression for the magnetic flux
\be
\Phi=2\pi N (-1+{\rm e}^{-e^2  F(1)}) .
\ee
The magnetic flux is not quantised, therefore these solutions, again, cannot exist on compact manifolds without boundary. They may, however, exist on non-compact manifolds (e.g., the punctured two-sphere) of on manifolds with boundary (e.g. a closed disc).

Note that $F(1)$ (and the flux) can be expressed as the target space averaged value of a function of the superpotential. Indeed,
\be
F(1)= \int_0^1 dh \frac{W(h)}{W_h (h)} = \left\langle \frac{W}{W_h}\right\rangle_{\mathbb{D}}
\ee
which is also valid for any $\mathcal{M}$ which supports solutions, at all.

\vspace*{0.2cm}

Now let us continue the computation of the volume of the soliton. The gauge field is
\be
1+a(y)={\rm e}^{e^2(F(h(y))-F(h=1))} .
\ee
This can be inserted into the first Bogomolny equation 
\be
dy =-N \frac{1+a}{W_h}dh=-N \frac{{\rm e}^{e^2(F(h)-F(1))}}{W_h} dh
\ee
Therefore, after integrating both sides we get
\be
V=2\pi y_0 = 2\pi N {\rm e}^{-e^2F(1)} \int_0^1 \frac{{\rm e}^{e^2 F(h)}}{W_h} dh .
\ee 
As the computation of the volume does not
require any particular form of solution, it is completely solution and coordinate independent (which means that also in the gauge case the volume is a thermodynamical function). Owing to the metric independence of the BPS equations it is, therefore, true for any manifold $\mathcal{M}$ with the right size and topology to host solutions. This allows us to write a Bradlow type relation for the existence of gauged {\em SDiff} vortices
\be
V_B \leq V' \leq V_{\mathcal{M}}
\ee
where the Bradlow volume is, once again, equal to the geometric volume of the soliton
\be
V_B=2\pi y_0 = 2\pi N {\rm e}^{-e^2 F(1)} \int_0^1 \frac{{\rm e}^{e^2 F(h)}}{W_h} dh .
\ee 
Let us also comment that the volume of the soliton can be written as an averaged integral
\be
V=2\pi N \left\langle \frac{ {\rm e}^{-4e^2 \left( \left\langle \frac{W}{W_h} \right\rangle_{\mathbb{D}} - \left\langle \frac{W}{W_h} \right\rangle_{\mathbb{D}_h} \right)}}{W_h} \right\rangle_{\mathbb{D}}
\ee
where $\mathbb{D}_h$ is a part of the unit disc parametrized by $(\tilde{h},\phi): \tilde{h}\in [0,h], \phi \in [0,2\pi]$.

To conclude, both non-gauged and gauged {\it SDiff} BPS models are very similar. For example, exactly as in the non-gauge case, if a potential supports infinitely extended $U(1)$ BPS vortices in flat space then, for any finite size manifold, the BPS sector will be empty. 

\section{Conformal solitons}
To complete our investigations we present solitonic models with conformal symmetry. 
As one could expect, the Bradlow volume takes zero value which means that the corresponding
BPS solitons can be constructed on an arbitrarily small compact manifold.

Let us consider the $CP^1$ model
\be
\mathcal{L}=\frac{u_\mu \bar{u}^\mu}{(1+|u|^2)^2}.
\ee
The BPS sector is defined by the Cauchy-Riemann condition 
\be
\bar{\partial} u =0\;\;\; \mbox{or} \;\;\;\partial u =0
\ee
solved by (anti)holomorphic functions. Such holomorphic functions can be defined 
also on a compact manifold, for example $\mathbb{S}^2_R$ with an arbitrary radius $R$ \cite{lump Sp1}. This can be 
achieved by use of the stereographic projection which relates coordinates on two sphere with usual complex variables $z \in \mathbb{C}$.  Since the action (and the Bogomolny equation) is conformally 
invariant the radius of the base space sphere can take any value. Therefore the corresponding Bradlow volume is
\be
V_B=0.
\ee
Note that the analysis is more subtle if the base space manifold is $T^2=\mathbb{S}^1 \times \mathbb{S}^1$. Then the BPS sector is not empty for $N>1$ and is given by the degree $N$ elliptic functions \cite{lump Sp2}. In the degree one case, it is possible to construct configurations with energy arbitrary close to $2\pi$ although equality is never attained. A similar situation occurs if we consider the Cauchy-Riemann equation on a disc. There are no non-constant holomorphic maps satisfying the obvious single point boundary condition. This means that the BPS sector is again empty.  

\vspace*{0.2cm}

The same concerns the instantons in the self-dual Yang-Mills $SU(2)$ model which can be constructed on $\mathbb{S}^4$ with arbitrary
radius. Another example of such conformal solitons (with $V_B=0$)  is provided by pure Skyrme instantons \cite{Sp3}. 
\section{Summary}

In Section II, we proved that the geometric volume of a soliton 
coincides with the thermodynamical volume also for models with local gauge symmetries. 
This identification holds for base space manifolds
whether they are flat or curved, compact or noncompact.

Then, using this geometric volume, we analysed the relationship between the soliton volume $V$, the volume of the manifold $V_\mathcal{M}$ and the Bradlow volume $V_B$, i.e., the minimal volume of the base space for which a given BPS theory  may possess a {\it non-empty} BPS sector. 

We found that the existence of a Bradlow bound seems to be a generic feature of BPS solitonic theories. By this we mean the following. Consider a solitonic field theory, which supports BPS solitons in flat (infinite) Minkowski space-time. Then consider the same action but on a manifold $\mathcal{M}$ with finite volume. There is a minimal volume of such a manifold (the Bradlow volume
 $V_B$) such that a solution to the (first-order) BPS equations can exist. 
If the volume of the manifold is further reduced, $V_\mathcal{M}<V_B$, then solutions cannot be BPS configurations but, rather, solitonic solutions of the full Euler-Lagrange equations. As a consequence, a non-zero pressure emerges as a quantity which characterises these solutions. All this shows that the Bradlow volume is a rather general concept which may be defined for many BPS models. In Section III.C.2 we found an exception to this rule, where the formal ``constant" solutions (\ref{const-sol}) have infinite energy, leading to an infinite formal Bradlow volume, whereas the model still supports compact BPS vortices of finite size. 
 
 \vspace*{0.2cm}
 We studied the Bradlow bound both for generalised abelian Higgs models and for a certain type of 
perfect fluid type models which we called {\it SDiff} BPS models (as the static energy functional is invariant under {\em SDiff} transformations of the base space), which in 1+1 dimensions reduce to the usual scalar field theory.
In particular, for the {\it SDiff} BPS models our findings can be summarised as follows. 
\begin{enumerate}
\item The Bradlow volume is equal to the geometric volume of the corresponding BPS soliton $V=V_B$. This says also that the geometric volume is base space independent. Here it is, of course, assumed that the base space manifold has the required minimal size and the right topology  to host  BPS solitons.
\item  A finite value of the Bradlow volume is observed if and only if the corresponding BPS solutions in flat Minkowski space-time are compactons. This, on the other hand, is completely controlled by the potential term or, more precisely, by its approach to the vacuum. The usual infinitely extended solitons (again in flat Minkowski space-time) give rise to an infinite value of the Bradlow volume. 
Note that such a simple relation between the type of solution (compacton/infinitely extended soliton) and the  value of the Bradlow volume (finite/infinite) is not a generic feature. Recall for example the Abelian Higgs model where solitons in $\mathbb{R}^2$ are infinitely extended although the Bradlow volume takes a finite value, leading to the relation $V \geq V_B$. 
\item BPS as well as non-BPS solitons are equivalent, whether one considers the model on a compact manifold with boundary or on a finite volume (non-compact) manifold. Therefore, there is no difference between {\it SDiff} vortices on $\mathbb{S}^2\backslash \{(0,0,-1)\}$ and $\mathbb{D} \subset \mathbb{R}^2$. As a consequence, reducing the volume of a manifold $\mathcal{M}$ is equivalent to squeezing the soliton into a smaller volume on the fixed background manifold. This allows for an easy access to the mean-field equation of state. Note again that this is different if compared to Abelian-Higgs vortices. 
\end{enumerate} 

Further, one can use the Bradlow volume to characterise general solitonic models. In fact, we found three types of theories with qualitatively different relations between the geometric volume of a soliton in the BPS sector and the Bradlow volume. 
\begin{enumerate}
\item Abelian Higgs type models - the BPS soliton always covers the full manifold $\mathcal{M}$ i.e., $V=V_\mathcal{M}$. Furthermore, the Bradlow bound requires $V \geq V_B$. The inequality is in fact saturated - BPS vortices exist for a compact manifold with $V_\mathcal{M} = V_B$
\item Perfect fluid ({\it SDiff}) models - $V=V_B$ i.e., the geometric (identical to thermodynamical) volume of the BPS soliton is base space (manifold $\mathcal{M}$) independent and obviously $V \leq V_\mathcal{M}$.
Some generalised abelian Higgs models supporting compactons behave in this way, as well.
\item Conformal models - again the BPS solitons cover the full manifold $\mathcal{M}$ i.e., $V=V_\mathcal{M}$. But now $V_B=0$. 
\end{enumerate}


It is interesting to note that the physical behaviour of the different phases is quite different between the abelian Higgs model and the {\em SDiff} models. In the {\em SDiff} models, the phase transition only exists for models supporting compactons, and the transition is from a fluid phase with constant nonzero pressure density but non-constant energy density for $V\le V_B$ to a gas of non-interacting compact vortices for $V>V_B$. For the abelian Higgs model (and for those generalisations which lead to non-compact BPS vortices), on the other hand, the phase transition is from the ``constant" solution (\ref{const-sol}) with constant energy density for $V\le V_B$ to a BPS solution which always covers the whole base space for $V\equiv V_\mathcal{M}>V_B$. The case of generalised abelian Higgs models supporting compact BPS vortices is somewhere in between, supporting the ``constant" solutions for $V\le V_B$, but supporting compact vortices (with a total volume equal to $V_B$), surrounded by empty space (vacuum) which fills the remaining volume $V_\mathcal{M} - V_B$. 

One very important step forward in this thermodynamical/fluid dynamical analysis of soliton models, obviously, is its generalisation to nonzero temperature. This generalisation is complicated by the fact that in a classical field theory there are infinitely many degrees of freedom which must be heated up (defrozen). In \cite{Manton1993} an approximate but exact equation of state at nonzero temperature $T$ was derived for the abelian Higgs model on a large sphere (corresponding to small $T$), by restricting the defreezing to the lightest degrees of freedom (the moduli).
A full thermodynamical description at nonzero temperature might well require the quantised version of the soliton model as a starting point. This endeavour, however, is very difficult owing to the perturbative non-renormalisability of said field theories. Only non-perturbative methods, therefore, have a chance to lead to interesting results.  

\section*{Acknowledgements}
CA acknowledges financial support from the Ministry of Education, Culture, and Sports, Spain (Grant No. FPA 2014-58-293-C2-1-P), the Xunta de Galicia (Grant No. INCITE09.296.035PR and Conselleria de Educacion), the Spanish Consolider-Ingenio 2010 Programme CPAN (CSD2007-00042), and FEDER.


\begin{thebibliography}{100}
\bb{Higgs1964}
P. Higgs, Phys. Lett. {\bf 12} (1964) 132.
\bb{Nielsen1973}
H.B. Nielsen, P. Olesen, Nucl. Phys. B{\bf 61} (1973) 45.
\bb{GL1950}
V.L. Ginzburg, L.D. Landau, Zh. Eksp. Teor. Fiz. {\bf 20} (1950) 1064.
\bb{Abrik1957}
A.A. Abrikosov, Sov. Phys. JETP {\bf 5} (1957) 1174.
\bb{Bogom1976}
E.B. Bogomolnyi, Sov. J. Nucl. Phys. {\bf 24} (1976) 449.
\bb{Prasad1975}
M.K. Prasad, C.M. Sommerfield, Phys. Rev. Lett. {\bf 35} (1975) 760.
\bb{Bradlow} S.B. Bradlow, Commun. Math. Phys. {\bf 135} (1990) 1.
\bb{volume} C. Adam, M. Haberichter, A. Wereszczynski, Phys. Lett. B{\bf 754} (2016) 18.
\bibitem{sp1}  J.M. Speight, J. Geom. Phys. {\bf 92} (2015) 30.
\bb{MS} N. Manton, P. Sutcliffe, Topological Solitons, Cambridge University Press, Cambridge, 2004.
\bb{Nasir} S.M. Nasir, Nonlinearity {\bf 11} (1998) 445.
\bb{five-vort}
N. Manton, 
 J. Phys. A {\bf 50} (2017) 125403.
\bb{Am-Ol}
 J. Ambjorn, P. Olesen,  Phys. Lett. B{\bf 214} (1988)
 565; Nucl. Phys. B{\bf 315} (1989)
 606.
\bb{Bja2017}
S.B. Gudnason, M. Nitta, arXiv:1701.04356. 
\bb{bazeia} D. Bazeia, E. da Hora, C. dos Santos, R. Menezes, Eur. Phys. J. C{\bf 71} (2011) 1833.
\bb{Lohe} M.A. Lohe, Phys. Rev. D{\bf 23} (1981) 2335.
\bb{contatto} F. Contatto, arXiv:1612.01879.
\bb{dun}
M. Dunajski, Phys. Lett. B{\bf 710} (2012) 236.
\bb{bazeia2016}
D. Bazeia, L. Losano, M.A. Marques, R. Menezes, I. Zafalan, 
arXiv:1611.02110. 

\bb{BPSgen} C. Adam, L.A. Ferreira, E. da Hora, A. Wereszczynski, W.J. Zakrzewski, JHEP{\bf 1308} (2013) 062.
\bb{BPSvortex} C. Adam, J. Sanchez-Guillen, A. Wereszczynski, W.J. Zakrzewski, Phys. Rev. D{\bf 86} (2013) 105009.
\bb{bps-baby}
C. Adam, T. Romanczukiewicz,  J. Sanchez-Guillen, A. Wereszczynski, 
Phys. Rev. D{\bf 81}, 085007 (2010).
\bibitem{GP}
T. Gisiger, M.B. Paranjape, Phys. Rev. D{\bf 55} (1997) 7731.
\bb{bps-baby Sp1}
J.M. Speight, 
J. Phys. A{\bf 43}, 405201 (2010). 
\bibitem{bps-baby Z} A.N. Leznov, B. Piette, W.J. Zakrzewski, J. Math. Phys. {\bf 38} (1997) 3007.
\bibitem{moser} J. Moser, Trans.\ Amer.\ Math.\ Soc. {\bf 120} (1965) 286.

\bibitem{hatcher} A. Hatcher, Algebraic Topology, Cambridge University Press, 2002.

\bb{BPSskyrme}
C. Adam, J. Sanchez-Guillen, A. Wereszczynski, 
Phys. Lett. B{\bf 691} (2010) 105.
\bb{gauged bps}
C. Adam, C. Naya, J. Sanchez-Guillen, A. Wereszczynski, Phys. Rev. D{\bf 86} (2012) 045010.
\bb{lump Sp1}  J.A. McGlade, J.M. Speight , Nonlinearity {\bf 19} (2006) 441; J.M. Speight,  J. Geom. Phys. {\bf 47} (2003) 343. 
\bb{lump Sp2} J.M. Speight,  Commun. Math. Phys. {\bf 194} (1998) 513.
\bb{Sp3} J.M. Speight,  Phys. Lett. B{\bf 659} (2008) 42.
\bb{Manton1993}
N. Manton,
Nucl. Phys. B{\bf 400} (1993) 624.


\end{thebibliography}
\end{document}